\documentclass[twocolumn,prb,showpacs]{revtex4}
\usepackage{graphicx}
\usepackage{dcolumn}
\usepackage{bm}
\begin{document}
\title{Checkerboard superconducting order and \\ antinodal
Bogoliubov quasiparticle interference}
\author{V.~I.~Belyavsky, V.~V.~Kapaev, and Yu.~V.~Kopaev}
\affiliation{P.~N.~Lebedev Physical Institute of Russian Academy
of Sciences, Moscow, 119991, Russia}%
\begin{abstract}
Numerical study of momentum-dependent gap function is presented to
make clear the origin of superconductivity in copper oxides. We
claim that antinodal region with pronounced nesting feature of the
Fermi contour gives rise to superconducting pairing with large
momentum under screened Coulomb repulsion. Such a pairing results
in both spatial checkerboard pattern of the superconducting state
below $T^{}_c$ and a gapped state of incoherent pairs in a broad
temperature range above $T^{}_c$. We explain the momentum
dependence of the coherent spectral weight detected in
angle-resolved photoemission spectroscopy and predict antinodal
Bogoliubov quasiparticle interference other than observed in the
nodal region.
\end{abstract}

\maketitle

\section{Introduction}

Angle resolved photoemission spectroscopy (ARPES) of underdoped
cuprates at temperatures exceeding conditional upper boundary
$T_{}^{\ast}$ of the pseudogap (PG) state evidences in favour to
large simply connected Fermi contour (FC) typical of the
conventional Fermi liquid. However, below $T_{}^{\ast}$, the FC is
seen as transformed into disconnected arcs disposed in the nodal
regions. Cooling from $T_{}^{\ast}$ down to superconducting (SC)
transition temperature $T^{}_c$ results in a decrease of arc
length down to zero. Thus, the FC degenerates into four points
that give rise to the nodes of $d$~-~wave SC order parameter
arising below $T^{}_c$. It seems fairly natural to conclude that,
within the framework of $d$~-~wave pairing concept, the SC order
parameter has its maximal value exactly in the antinodal
directions.\cite{Norman}

For this reason, it might seem quite probable that low-temperature
properties of $d$~-~wave superconductor should be determined by
low-energy quasiparticle excitations only in the nodal region of
the momentum space that is in vicinities of the SC gap nodes on
the diagonals of the Brillouin zone. Taking into account that
Bogoliubov quasipatricle interference (QPI), observed in the nodal
region, disappears near the end points of the FC
arcs,\cite{Kohsaka} one might lead to a conclusion that only the
nodal region gives rise to superconductivity whereas the gap
observed in the antinodal region should be attributed to an
incoherent PG state.\cite{Kohsaka} However, a coherence in the
antinodal region becomes apparent both in the ARPES study
\cite{Kondo} and also in the Andreev~--~St~James
experiments.\cite{Deutscher} Therefore, in spite of the fact that
high-energy QPI is not detected for the present, one can believe
that the antinodal region should contribute a coherent state as
well.

We have argued\cite{BK} that both PG and SC states arise exactly
in the antinodal region with pronounced nesting of the FC as
spatially inhomogeneous incoherent and coherent states of pairs
with large momentum, respectively. The nodal region gives rise to
conventional SC pairing with zero momentum which, together with
the pairing with large momentum (${\bm{K}}$-pairing) in the
antinodal region, forms a {\it{biordered}} SC state in the whole
of the Brillouin zone.

Kinematic constraint, inherent in ${\bm{K}}$-pairing in the
antinodal region, can result in oscillating real-space pairing
interaction. Indeed, momenta of both particles composing SC pair
with nonzero total momentum ${\bm{K}}$ should be either inside or
outside of the FC. For this reason, a set of one-particle states
turns out to be kinematically excluded because of the fact that
such states cannot contribute into the states of ${\bm{K}}$-pairs.
It means that any scattering between such excluded states should
be forbidden when one defines the interaction leading to a rise of
a bound state of ${\bm{K}}$-pair. An exclusion of a set of the
Fourier components from the screened Coulomb interaction results
in the fact that corresponding real-space  ${\bm{K}}$-pairing
interaction exhibits an oscillation outside of small-distance
repulsive core as shown schematically in Fig.~1. It should be
noted that there is an analogy between this oscillation and
well-known Friedel oscillation that arises owing to Kohn
singularity of screening enhanced by nesting of the FC.

Besides the fact that two-particle problem with oscillating
potential leads to a bound state of the relative motion of
${\bm{K}}$-pair,\cite{BKS} it can also produce a quasi-stationary
state (QSS)\cite{BKTS} similar to the Gamov's state of
alpha-radioactive nucleus.\cite{Gamov} SC gap function
${\Delta}^{}_{sc}({\bm{k}})$, depending on relative-motion
momentum ${\bm{k}}$ of ${\bm{K}}$-pair, as a solution to the
mean-field self-consistency equation, arises as a result of the
instability of the ground state of the normal Fermi liquid with
respect to a rise of ${\bm{K}}$-pairs in the bound state. This
function can be expressed in terms of Gorkov's anomalous averages
describing SC condensate of ${\bm{K}}$-pairs. Due to a phase
coherence of the SC ground state, these averages become nonzero
below $T^{}_c$. QSS with positive energy, following from the
two-particle problem, can be considered as an evidence in favor to
one more instability of the Fermi liquid. We associate such an
instability with incoherent ${\bm{K}}$-pairs existing above
$T^{}_c$. It means that Gorkov's anomalous averages and
corresponding gap function ${\Delta}^{}_{pg}({\bm{k}})$ vanish
under averaging over phases of ${\bm{K}}$-pairs but mean square
gap function remains nonzero up to $T^{\ast}_{}$ according to the
hypothesis advanced by Emery and Kivelson.\cite{Emery}

\begin{figure}
\includegraphics[]{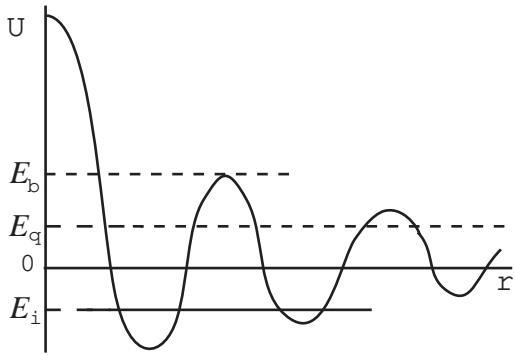}
\caption[*]{Real-space pairing potential $U(r)$ (schematically).
Energies $E^{}_i$ and $E^{}_q$ correspond to bound and
quasi-stationary states, respectively. Barrier height $E^{}_b$
corresponds to a break of the pair without tunnelling through the
barrier.}\label{F2.eps}
\end{figure}

Thus, one can conclude that, in the temperature range from
$T_{}^{\ast}$ down to $T^{}_c$, ${\bm{K}}$-pairs form incoherent
PG state as off-diagonal short-range order (ODSRO). Off-diagonal
long-range order (ODLRO) arises as SC condensate of coherent
${\bm{K}}$-pairs below $T^{}_c$. Both ODLRO and ODSRO states can
be described in terms of Gorkov's Green functions.\cite{AGD} A
phenomenological BCS-like form of the coherent contribution to the
normal Gorkov's function can be written as
$$
G^{}_{}({\omega};{\bm{k}})=z({\bm{k}}) \left
[{\frac{u^{2}_+({\bm{k}})}
{{\omega}-E({\bm{k}})+i{\Gamma}}}+{\frac{ u^{2}_-({\bm{k}})}
{{\omega}+E({\bm{k}})-i{\Gamma}}}\right ],
$$
where $E({\bm{k}})$ and $2u^{2}_{\pm}({\bm{k}})=1\pm
{\xi}({\bm{k}})/E({\bm{k}})$ are quasiparticle energy and
coherence factors, respectively,
\begin{equation}\label{0}
2{\xi}({\bm{k}})={\varepsilon}({\bm{K}}/2+{\bm{k}})+
{\varepsilon}({\bm{K}}/2-{\bm{k}})
\end{equation}
is the kinetic energy of the ${\bm{K}}$~-~pair of particles with
momenta ${\bm{K}}/2\pm {\bm{k}}$, ${\varepsilon}({\bm{k}})$ is
electron dispersion with respect to chemical potential ${\mu}$ and
$z({\bm{k}})$ is quasiparticle weight. Two terms in
$G^{}_{}({\omega};{\bm{k}})$ can be referred to ${\bm{K}}$-pairs
above and below the FC, respectively. Diagonal Green function
$G^{}_{}({\omega};{\bm{k}})$ describes ODSRO state corresponding
to the existence of non-coherent QSS of ${\bm{K}}$~-~pairs above
$T^{}_c$. Transition from the bound paired state into long-living
QSS corresponds to small but finite decay
${\Gamma}={\Gamma}({\omega};{\bm{k}})$ whereas transitions into
stationary states above barrier energy $E^{}_b$ (Fig.~1) should be
associated with an infinitesimal decay, ${\gamma}\rightarrow +0$,
leading to conventional Fermi-liquid behavior of
$G^{}_{}({\omega};{\bm{k}})$ above $T^{\ast}_{}$. Thus, a rise of
QSS results in a non-Fermi-liquid behavior of
$G^{}_{}({\omega};{\bm{k}})$) that can be related to the PG state.

The SC state below $T^{}_c$ should be described by both normal and
anomalous Gorkov's functions. Taking into account the fact that PG
function ${\Delta}^{}_{pg}({\bm{k}})$, averaged over random
phases, vanishes whereas ${\Delta}^{}_{sc}({\bm{k}})\neq 0$ below
$T^{}_c$, one can introduce anomalous Gorkov's function
$F_{}^+({\omega};{\bm{k}})$ in a way we use to obtain
$G^{}_{}({\omega};{\bm{k}})$:
$$
F_{}^+({\omega};{\bm{k}})=-z({\bm{k}}){\frac{
{\Delta}^{\ast}_{sc}({\bm{k}})}
{{({\omega}-E({\bm{k}})+i{\Gamma})}
{({\omega}+E({\bm{k}})-i{\Gamma})}}}.
$$

Such an approach directly leads to uniform description of both SC
and PG states in underdoped cuprates. One can see that repulsive
Coulomb pairing in the antinodal region necessarily results in
rather complicated momentum dependence of the SC gap and PG
functions, ${\Delta}^{}_{sc}({\bm{k}})$ and
${\Delta}^{}_{pg}({\bm{k}})$, with energy scale
${\varepsilon}^{}_0\sim 1\, eV$ of their domains of
definition\cite{BKK} in contrast to considerably less scale of
about Debye energy ${\varepsilon}^{}_D$ that arises in the case of
phonon-mediated SC pairing. It is very likely that the {\it{high
energy problem}},\cite{Leggett} arising, in particular, in the
optical conductivity of the cuprates,\cite{Basov} might be
associated with high energy scale of the antinodal
${\bm{K}}$-pairing. We believe that ${\bm{K}}$-pairs are the main
players in the high-temperature superconductivity of the cuprates.

Recently, Tsvelik and Chubukov\cite{Tsvelik} considered SC pairing
on mutually orthogonal pairs of perfectly nested segments of the
FC in semiphenomenological way. They presuppose that
one-dimensional SC order arises only on these segments coupled
with a momentum-space Josephson links to give rise to
two-dimensional superconductivity. Actually, it is implicitly
supposed that such SC state can arise owing to SC pairing with
nonzero momentum. Also, it is supposed that SC order with the same
momentum is induced on the rest unnested part of the FC by the
order on the nested segments due to a {\it{proximity effect in the
momentum space}}\cite{BKS} so that this induced order cannot
penetrate deep into the nodal region. It should be emphasize that
such a model\cite{Tsvelik} differs essentially from the biordered
SC state.\cite{BK}

In this paper, we study the mean-field ${\bm{K}}$-pairing problem
numerically to fall outside the weak coupling limits employed in
our previous approach to the ${\bm{K}}$-pairing
problem.\cite{BK,BKS,BKTS} We show that the SC gap function with a
nontrivial nodal line corresponds to a checkerboard {\it{pair
density wave}} (PDW) SC state and results in fairly natural
explanation of the angle dependence of a partial suppression of
the coherent spectral weight in the antinodal region observed by
Kondo et al.\cite{Kondo} We believe that QPI, other than observed
in the nodal region,\cite{Kohsaka} could be detected in the
antinodal one as well. We also show that ${\bm{K}}$-pairing can
originate spatial checkerboard pattern without any driving
insulating order in contrast to a scenario of a rise of PDW
coexisting with a charge density wave (CDW).\cite{Chen}


\section{$K$~-~pairing problem}

In the case of ${\bm{K}}$-pairing, the gap function is defined as
\begin{equation}\label{1}
{\Delta}({\bm{k}})=\sum_{{\bm{k}}_{}^{\prime}}
U({\bm{k}},{\bm{k}}_{}^{\prime})\langle
{\hat{c}}^{}_{{\bm{K}}/2-{\bm{k}}_{}^{\prime}\downarrow}
{\hat{c}}^{}_{{\bm{K}}/2+{\bm{k}}_{}^{\prime}\uparrow} \rangle
\end{equation}
where $U({\bm{k}},{\bm{k}}_{}^{\prime})$ is screened Coulomb
interaction matrix element, operator
${\hat{c}}^{}_{{\bm{k}\sigma}}$ annihilates electron with momentum
${\bm{k}}$ and spin polarization $\sigma$. Anomalous average in
Eq.~(\ref{1}), describing SC condensate of ${\bm{K}}$~-~pairs,
becomes nonzero below $T^{}_c$. The gap function should be a
nontrivial solution to the self-consistency equation,
\begin{equation}\label{2}
{\Delta}({\bm{k}})=-{\frac{1}{2}}\sum_{{\bm{k}}_{}^{\prime}}
{\frac{U({\bm{k}},{\bm{k}}_{}^{\prime}){\Delta}({\bm{k}}_{}^{\prime})}
{{\sqrt{{\xi}^2_{}({{\bm{k}}_{}^{\prime}})+{\Delta}^2_{}({\bm{k}}_{}^{\prime})}}}}
[1-n({{\bm{k}}_{}^{\prime}})].
\end{equation}
Here, $n({{\bm{k}}})=(e_{}^{E({\bm{k}})/T}+1)_{}^{-1}$ is a
quasiparticle occupation number and quasiparticle energy has the
form
\begin{equation}\label{B}
E({\bm{k}})={\eta}({\bm{k}})\pm {\sqrt{{\xi}_{}^2({\bm{k}})+
{\Delta}^2_{}({\bm{k}})}}
\end{equation}
\begin{equation}\label{C}
2{\eta}({\bm{k}})={\varepsilon}({\bm{K}}/2+{\bm{k}})-
{\varepsilon}({\bm{K}}/2-{\bm{k}}).
\end{equation}
It should be noted that, since ${\varepsilon}(-{\bm{k}})=
{\varepsilon}({\bm{k}})$ owing to the time-reversal symmetry of
the dispersion relation, quasiparticle spectrum (\ref{B}) turns
out to be gapped on the whole of the FC in the case of pairing
with zero total momentum. In the case of ${\bm{K}}$-pairing, it
can be gapped only on those parts of the FC where
$|{\eta}({\bm{k}})|$ proves to be small enough. In addition, the
quasiparticle spectrum becomes asymmetrical with respect to
${\mu}$.

Summation in Eqs.~(\ref{1}) and (\ref{2}) should be performed over
all momenta of the relative motion which can form pairs with given
total momentum ${\bm{K}}$. One can see that these momenta belong
to a ${\bm{K}}$~-~dependent domain of the momentum space (domain
of {\it{kinematic constraint}}) because of the fact that the
momenta of both particles composing SC pair with given total
momentum should be situated either inside or outside of the FC.
This means that some part of the momentum space turns out to be
excluded from the sums in Eqs.~(\ref{1},\ref{2}).

Since the kinetic energies of both particles composing SC pair
with ${\bm{K}}=0$ can be equal to ${\mu}$, the low-energy limit in
the sum (\ref{2}) corresponds to ${\xi}=0$ whereas the upper limit
is formally restricted by a half-width of the conduction band of
the order of ${\mu}$. However, in the Bardeen-Cooper-Schrieffer
(BCS) theory,\cite{BCS} such upper limit (Debye phonon energy
${\varepsilon}^{}_D$) appears as an energy scale of a layer
enveloping the FC where electron-electron scattering results in an
effective attraction between electrons. As a result, pairing
interaction energy $U({\bm{k}},{\bm{k}}_{}^{\prime})$ in
Eq.~(\ref{2}) can be qualitatively associated with an effective
coupling constant $V_{}^{\ast}$ that can be estimated
as\cite{Bogoliubov}
\begin{equation}\label{D}
V_{}^{\ast} =V-{\frac{U}{1+Ug{\ln{({\mu}/{\varepsilon}^{}_D)}}}},
\end{equation}
where $g$ is density of states per spin, $V$ is a pairing constant
due to electron-phonon interaction defined inside the layer, $U$
is average Coulomb energy. Thus, in the effective pairing
constant, Coulomb repulsion appears with a logarithmic weakening.
In the case when $V_{}^{\ast}g\ll 1$, the mean-field approach
results in a conventional BCS energy gap,
\begin{equation}\label{E}
{\Delta}=2{\varepsilon}^{}_D {\exp{(-1/V_{}^{\ast}g)}},
\end{equation}
that appears in consequence of the logarithmic singularity of the
right-hand side of Eq.~(\ref{2}). This singularity is primarily
formed in an energy range near the low-energy limit, therefore,
extension of this range might lead to a progressive accumulation
of the singularity along with the formation of a non-singular
(regular) contribution into Eq.~(\ref{2}). One can treat the
preexponential in Eq.~(\ref{E}) as a characteristic energy scale
beyond which the nontrivial solution to the self-consistency
equation becomes weakly sensitive to the upper limit.

\begin{figure}
\includegraphics[scale=.36]{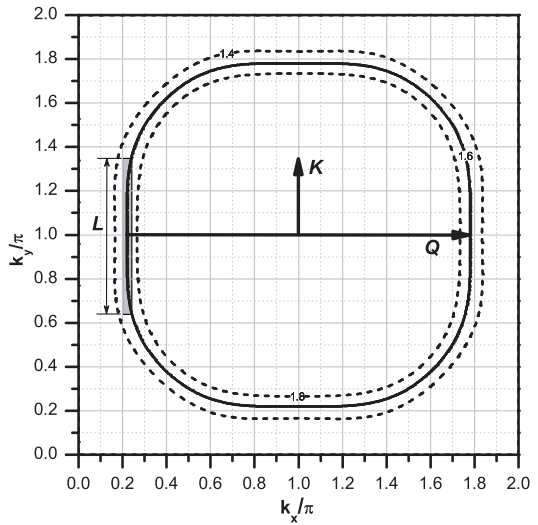}
\caption[*]{Near nested FC (solid line) corresponding to electron
dispersion Eq.~(\ref{4}) typical of the cuprates. Dashed lines
represent isolines close to the FC, numbers near the isolines are
electron energies according to Eq.~({\ref{4}}) with $t=0.5$,
$t_{}^{\prime}=-0.15$, $t_{}^{\prime \prime}=0.07$ $eV$. Here,
${\bm{K}}$ is total momentum of ${\bm{K}}$-pair, ${\bm{Q}}$ is
nesting momentum. Length $L$ of near rectilinear segment of the FC
is shown at given mean square energy deviation ${\delta}$ (the
width of the shadowed strip) of the FC from the
rectilinearity.}\label{FC.eps}
\end{figure}

All these speculations can be referred to the ${\bm{K}}$~-~pairing
problem. However, in such a case, the logarithmic singularity
becomes apparent if and only if kinetic energy of
${\bm{K}}$~-~pair vanishes not at isolated points, as it were most
likely in the case of arbitrary FC, but on finite pieces of the FC
on which {\it{mirror nesting}} condition,
\begin{equation}\label{3}
{\varepsilon}({\bm{K}}/2+{\bm{k}})-
{\varepsilon}({\bm{K}}/2-{\bm{k}})=0,
\end{equation}
should be fulfilled at given ${\bm{K}}$. One can see that, for a
suitable ${\bm{K}}$, this condition can be fulfilled in the case
of rectilinear parallel segments on the opposite sides of the FC.
It is obvious that ${\bm{K}}$ should be directed along these
segments.

FC, typical of the cuprates, and isolines close to it can be
described satisfactorily by electron dispersion
\begin{eqnarray}\label{4}
{\varepsilon}(k^{}_x,k^{}_y)&=&t^{}_0-2t({\cos{k^{}_x}}+{\cos{k^{}_y}})
-4t_{}^{\prime}{\cos{k^{}_x}}{\cos{k^{}_y}}-\nonumber \\
&-&2t_{}^{{\prime}{\prime}}({\cos{2k^{}_x}}+{\cos{2k^{}_y}})
\end{eqnarray}
with fitting parameters $t^{}_0=2\,e\, V$ $t=0.5\,e\, V$,
$t_{}^{\prime}/t=-0.3$, $t_{}^{{\prime}{\prime}}=0.14$. Here,
$k^{}_x$ and $k^{}_y$ (in units of ${\pi}/a$; $a$ is interatomic
distance) are momentum components corresponding to the antinodal
directions.

One can choose one of the coordinate axes ($k^{}_y$) along
${\bm{K}}$ directed parallel to antinodal near rectilinear
segments of the FC as shown in Fig.~2. Then, with preassigned
accuracy ${\delta}$, ${\xi}({\bm{k}})\leq {\delta}$ if $k^{}_y$
corresponds to near rectilinear segment of the FC. Therefore, the
singular contribution into (\ref{2}) turns out to be proportional
to the length $L$ of such a segment. Summation over the other
component ($k^{}_x$) leads to an accumulation of the singularity,
however, in contrast to the case ${\bm{K}}=0$, a gradual deviation
from the FC results in a progressive increase of the difference
between the kinetic energies of the particles composing
${\bm{K}}$~-~pair. This leads to increasing deviation from mirror
nesting condition (\ref{3}), so that, finally, the accumulation
turns out to be completed when $k^{}_x$ attains a value
corresponding to energy scale ${\varepsilon}^{}_0$ much lesser
than ${\mu}$. It should be noted that, in the case of
${\bm{K}}$~-~pairing, ${\varepsilon}^{}_0$ appears as generic
energy scale originating from mirror nesting feature of electron
dispersion. This scale should be related to a preexponential of
the gap function in the case of small effective coupling constant.
Thus, one can conclude that the nontrivial solution to the
self-consistency equation should be weakly sensitive to the part
of the momentum space corresponding to
${\varepsilon}>{\varepsilon}^{}_0$.

\begin{figure}
\includegraphics[scale=.32]{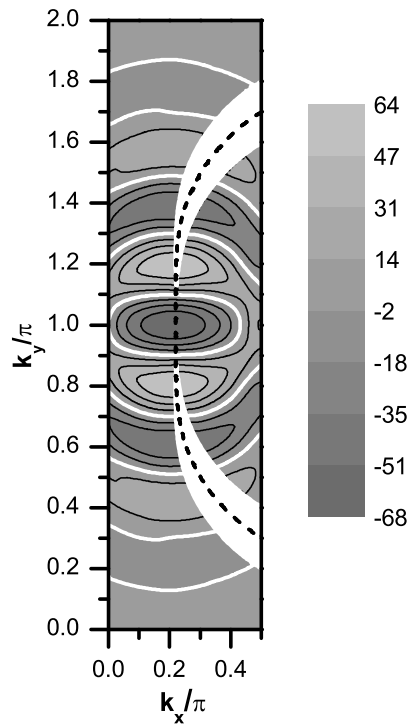}
\caption[*]{Calculated momentum dependence of the antinodal gap
function (gradation of grey, in $meV$) shown in a region of the
Brillouin zone that primarily forms the singularity of the
self-consistency equation. White curves represent the nodal line
structure of the gap function, dashed line is the FC corresponding
to that shown in Fig.~2. White regions, in which the nontrivial
solution is absent due to the kinematic constraint, appear under
shifting of the FC at $\pm{\bm{K}}/2$ along $k^{}_y$ (inside these
regions, there are no one-particle states that could form a pair
with total momentum $K=0.2\,{\pi}/a$) corresponding to a maximum
magnitude of the gap function.}\label{FC.eps}
\end{figure}

Since equality (\ref{3}) is fulfilled only approximately, one can
choose length $L$ in order that mean square deviation of the FC
from the rectilinearity were less than a preassigned value
corresponding to energy scale ${\delta}$. Strictly speaking,
nonzero ${\delta}$ eliminates the singularity because of a rise of
a lower limit cutoff in the sum Eq.~(\ref{2}). Similar cutoff
appears in the Fulde~-~Ferrell~-~Larkin~-~Ovchinnikov (FFLO)
problem of SC pairing with small total pair momentum.\cite{FF,LO}
Therefore, nontrivial solution to equation (\ref{2}) can exist if
the effective coupling constant exceeds certain
${\delta}$~-~dependent value. Magnitude ${\Delta}_{}^{\prime}$ of
the gap function can be roughly estimated as
\begin{equation}\label{5}
{\Delta}_{}^{\prime}={\sqrt{{\Delta}({\Delta}-2{\delta})}},
\end{equation}
where ${\Delta}$ is the magnitude corresponding to perfect mirror
nesting that is to exactly rectilinear segment of the FC with
length $L$. Positive function ${\Delta}_{}^{\prime}({\delta})$ has
a maximum at certain ${\delta}\equiv {\delta}_{m}^{}< {\Delta}/2$.
Indeed, ${\Delta}_{}^{\prime}\rightarrow 0$ if
${\delta}\rightarrow 0$ (then, generally speaking, $L\rightarrow
0$, so that there is no singularity in the self-consistency
equation: ${\Delta}\rightarrow 0$). At $2{\delta}> {\Delta}$, the
magnitude of the gap function vanishes, therefore, a maximum value
of ${\Delta}_{}^{\prime}$ exists at
$0<{\delta}_{m}^{}<{\Delta}/2$. A choice of length $L$ of near
rectilinear segment of the FC at given ${\delta}$ predetermines
total pair momentum ${\bm{K}}$. It is clear that, because of
kinematic constraint, the absolute value of ${\bm{K}}$ coincides
with $L/2$, as one can see from Fig.~2. Since maximum value of
${\Delta}_{}^{\prime}$ corresponds to ${\delta}^{}_m$, the
absolute value of the momentum of ${\bm{K}}$~-~pairs in the SC
condensate should be taken as $K=L({\delta}^{}_m)/2$. Variation of
the FC with doping $x$ in hole doped cuprates\cite{Fujimori}
results in a conclusion that $K$ should be dependent on $x$.  Note
that there is no contradiction between such a dependence,
following from dispersion (\ref{4}), and doping dependence of
spatial periodicity of checkerboard PDW seen in tunnel
data.\cite{Wise}

Comparatively small vicinity with energy scale
${\varepsilon}^{}_0$ of the strip with length $L({\delta}^{}_m)/2$
and width corresponding to energy scale ${\delta}^{}_m$ can be
considered as the region of the momentum space that primarily
forms the singularity of the self-consistency equation. Following
Ref.~[10], one can renormalize the kernel of this equation and
reduce Eq.~(\ref{2}) to a sum over momenta belonging to such a
vicinity only. Renormalized kernel, defined in this vicinity, can
be written as\cite{BKNT}
\begin{equation}\label{F}
W({\bm{k}},{\bm{k}}_{}^{\prime})=\sum_n
{\frac{{\phi}^{}_n({\bm{k}}){\phi}^{\ast}_n({\bm{k}}_{}^{\prime})}
{{\lambda}^{}_n+g\,{\ln{({\mu}/{\varepsilon}^{}_0)}}}},
\end{equation}
and can be treated as a pairing pseudopotential corresponding to
oscillating real-space pairing interaction. Here,
${\phi}^{}_n({\bm{k}})$ and ${\lambda}^{}_n$ are eigenfunctions
and eigenvalues of kernel $U({\bm{k}},{\bm{k}}_{}^{\prime})$,
respectively.\cite{BKS} We believe that the vicinities with energy
scale ${\varepsilon}^{}_0$ of the antinodal near rectilinear
segments of the FC include electron states that mainly contribute
to scattering resulting in ${\bm{K}}$-pairing.

Since characteristic sizes of the vicinity are much less than
characteristic Fermi momentum, region of attraction in the real
space proves to be more deep and extended with respect to that due
to Friedel oscillation. Such oscillating interaction can provide
both bound state and QSS of the relative motion of
${\bm{K}}$~-~pair. In the mean-field approach, the bound state
appears in temperature range $0\leq T<T^{}_c$ as nonzero anomalous
averages, $\langle {\hat{c}}^{}_{{\bm{K}}/2-{\bm{k}}\downarrow}
{\hat{c}}^{}_{{\bm{K}}/2+{\bm{k}}\uparrow} \rangle \neq 0$, that
determine gap function Eq.~(\ref{1}). It should be noted that, in
the case of small ${\bm{K}}$ (for example, in the FFLO state),
real-space oscillation of the pairing interaction becomes weak
enough because of considerable extension of the corresponding
vicinity forming the singular contribution into the
self-consistency equation.

To study ${\bm{K}}$-pairing problem numerically, we use a
step-wise approximation of the pairing interaction\cite{BKK}
assuming that pseudopotential (\ref{F}) has a constant value of
about $10\; eV$ inside a vicinity of near rectilinear segments of
the FC. Energy scale of such a vicinity is determined from the
above mentioned condition that calculated gap function magnitude
should become actually independent of this scale beginning with
certain ${\varepsilon}^{}_0$.

Numerical study of Eq.~(\ref{2}) at $T=0$ reveals highly
complicated momentum dependence of gap function
${\Delta}({\bm{k}})$, shown in Fig.~3, with a few closed nodal
lines crossing the FC. Topological feature of the gap function,
shown only inside the part of the Brillouin zone that primarily
contributes into the singularity of the self-consistency equation,
turns out to be weakly dependent on small variation of the
parameters of electron dispersion and magnitude of pairing
interaction. According to rough estimation following from
(\ref{5}), maximum value of $K$~-~dependent magnitude of the gap
function can be associated with $K$ close to $0.2\, {\pi}/a$.
Domain of definition of the pairing pseudopotential includes all
energies ${\varepsilon}<{\varepsilon}^{}_0$, where
${\varepsilon}^{}_0$ is relative to a distance between the FC and
the boundary of this domain. As follows from numerical solution to
Eq.~(\ref{2}), a gradual decrease of the momentum corresponding to
upper limit $k^{}_r$ in the sum over $k^{}_x$ in the
self-consistency equation with renormalized kernel, at first, does
not affect the magnitude of the gap function. Then, beginning with
certain value of $k^{}_r$, that can be associated with a boundary
of the domain of definition of $W({\bm{k}},{\bm{k}}_{}^{\prime})$,
the magnitude tends to zero with a decrease of $k^{}_r$. This
gives a possibility to determine energy scale
${\varepsilon}^{}_0\approx 0.3\,eV$ of this domain forming the
singularity of the self-consistency equation.

\section{Checkerboard ODLRO}

Visualization of a checkerboard PDW\cite{Wise} can be considered
as an indirect evidence in favour of the fact that such a state
originates from nesting feature of the FC typical of the cuprates.
It should be noted that near rectilinear segments on the opposite
sides of such FC ensure not only mirror nesting condition
(\ref{3}) but also nesting condition
\begin{equation}\label{6}
{\varepsilon}({\bm{k}}+{\bm{Q}})+ {\varepsilon}({\bm{k}})=0
\end{equation}
at certain nesting momentum ${\bm{Q}}$ which, in general, is
incommensurate with total momentum ${\bm{K}}$ of SC pair as shown
in Fig.~2. Under condition (\ref{6}), the logarithmic singularity
can arise in an insulating pairing channel that gives rise, for
example, to CDW. Such an insulating pairing can compete or coexist
with ${\bm{K}}$~-~pairing in a way considered a long time ago in
the case of the coexistence of conventional ($K=0$) SC state and
CDW.\cite{Rusinov}

One can compare efficiencies of both channels with the help of a
crude estimation of lengths $L^{}_K$ and $L^{}_Q$ of near
rectilinear segments forming singularities in the SC and
insulating channels, respectively. These lengths, at given
${\delta}$, the same in both channels, can be found from
inequalities
\begin{eqnarray}\label{G}
|{\varepsilon}({\bm{K}}/2+{\bm{k}})-
{\varepsilon}({\bm{K}}/2-{\bm{k}})|\leq {\delta}, \nonumber \\
|{\varepsilon}({\bm{k}}+{\bm{Q}})+ {\varepsilon}({\bm{k}})| \leq
{\delta},
\end{eqnarray}
selecting the regions in the momentum space in which mirror
nesting or nesting condition, respectively, is satisfied with
preassigned accuracy. If boundaries of these regions intersect the
FC, lengths $L^{}_K$ and $L^{}_Q$ should be defined as distances
between the corresponding intersection points. Both pair momentum
${\bm{K}}$ and nesting momentum ${\bm{Q}}$ should be selected in a
way to ensure maximum values of corresponding lengths $L^{}_K$ and
$L^{}_Q$, respectively. Momenta ${\bm{K}}$ and ${\bm{Q}}$ depend
on a form of the FC varying with doping. Therefore, interrelation
between $L^{}_K$ and $L^{}_Q$ varies with doping as well.
Calculated variations of $L^{}_K$ and $L^{}_Q$ with doping are
shown schematically in Fig.~4. A comparison of $L$ and $L^{}_Q$
shows that, in the case of electron dispersion Eq.~(\ref{4}),
nesting dominates mirror nesting in electron doped compounds. On
the contrary, the opposite case of hole doping gives an
opportunity of a rise of such a range of $x$ where mirror nesting
dominates nesting.

\begin{figure}
\includegraphics[scale=.36]{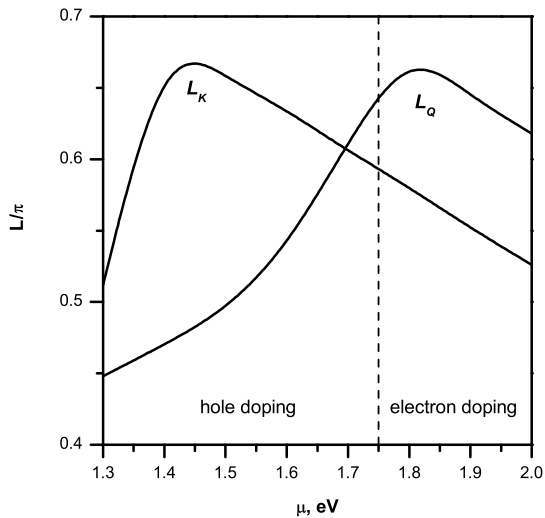}
\caption[*]{Competition between nesting and mirror nesting: doping
dependence of lengths $L^{}_Q$ and $L$ of the optimal near
rectilinear segments of the FC under nesting and mirror nesting
conditions, respectively.}\label{FC.eps}
\end{figure}

In this range, SC order arises due to ${\bm{K}}$~-~pairing whereas
insulating order (CDW with $Q=2K$) with low spectral weight can
exist as induced by the corresponding PDW and, therefore, turns
out to be hardly detected. Conversely, dominating insulating order
in the form of CDW can induce the PDW\cite{Shen} as a modulated
superfluid density which, due to low spectral weight, seems to be
undetectable. Thus, one can conclude that, if mirror nesting
dominates nesting, SC state in the form of PDW of
${\bm{K}}$~-~pairs can exist without any driving insulating order.

It should be emphasized that nesting feature of the FC can lead to
a giant enhancement of the singularity in both pairing
channels.\cite{Rusinov,Furukawa} In the insulating channel,
nesting of the FC transforms weak Kohn anomaly into the CDW. In
the SC channel, mirror nesting gives rise to ${\bm{K}}$-pairing
which, owing to kinematic constraint, ensures extended and deep
oscillation of real-space screened Coulomb pairing interaction
and, consequently, results in $T^{}_c$ considerably greater than
following from Kohn-Luttinger SC pairing\cite{KL} with
${\bm{K}}=0$ and angular momentum $l\neq 0$.

Thus, ${\bm{K}}$-pairing leads to an independent order so that, in
such a case, there is no need to take into account a coexistence
of SC and insulating ordered states\cite{Chen} to invoke spatially
inhomogeneous SC state in the form of PDW.

One can define anomalous averages $\langle
{\hat{\psi}}^{}_{\downarrow}({\bm{r}}_{}^{\prime})
{\hat{\psi}}^{}_{\uparrow}({\bm{r}})\rangle $ in the real space
that corresponds to the momentum-space anomalous averages,
$\langle {\hat{c}}^{}_{{\bm{K}}/2-{\bm{k}}\downarrow}
{\hat{c}}^{}_{{\bm{K}}/2+{\bm{k}}\uparrow}\rangle $, arising due
to ${\bm{K}}$-pairing. Here, fermion field operator
${\hat{\psi}}^{}_{\sigma}({\bm{r}}) $ annihilates electron with
spin polarization $\sigma$ and radius vector ${\bm{r}}$. Nonzero
anomalous averages can be considered as an order parameter
corresponding to ${\bm{K}}$-pairing. In the case of the
two-dimensional $C^{}_4$ orbital symmetry, there are four crystal
equivalent pair momenta ${\bm{K}}^{}_j$, $j=1,2,3,4$. Therefore,
real-space representation of the order parameter should be written
as a superposition
\begin{equation}\label{7}
\langle {\hat{\psi}}^{}_{\downarrow}({\bm{r}}_{}^{\prime})
{\hat{\psi}}^{}_{\uparrow}({\bm{r}})\rangle =\sum_{j=1}^4
{\gamma}^{}_j\, {\varphi}^{}_j({\bm{\rho}})\cdot
{\exp{(i{\bm{K}}^{}_j{\bm{R}})}},
\end{equation}
where ${\bm{R}}=({\bm{r}}+{\bm{r}}_{}^{\prime})/2$, ${\bm{\rho}}=
{\bm{r}}-{\bm{r}}_{}^{\prime}$ are center-of-mass and relative
motion radius vectors of ${\bm{K}}$-pair,
\begin{equation}\label{8}
{\varphi}^{}_j({\bm{\rho}})={\frac{1}{N}}\sum_{\bm{k}} \langle
{\hat{c}}^{}_{{\bm{K}}^{}_j/2-{\bm{k}}\downarrow}
{\hat{c}}^{}_{{\bm{K}}^{}_j/2+{\bm{k}}\uparrow}\rangle \,
{\exp{(i{\bm{k}}{\bm{\rho}})}}
\end{equation}
can be considered as a real-space wave function of the relative
motion of ${\bm{K}}$-pair. Here, $N$ is a number of unit cells of
the two-dimensional system, summation over ${\bm{k}}$ should be
performed inside the domain of kinematic constraint corresponding
to each momentum ${\bm{K}}^{}_j$. Coefficients ${\gamma}^{}_j$,
corresponding to SC state, should be determined by one of the
irreducible representations of the symmetry group $C^{}_4$. A
choice of the irreducible representation establishes the orbital
symmetry of the order parameter. Since ${\gamma}^{}_1=
-{\gamma}^{}_2= {\gamma}^{}_3= -{\gamma}^{}_4$ in the case of
$d$-wave orbital symmetry, a checkerboard spatial pattern of the
order parameter follows from Eq.~(\ref{7}) immediately. One can
see that $d$-wave order parameter (\ref{7}) corresponds to a
current-less SC state, therefore, in this respect, it is similar
to Larkin~-~Ovchinnikov immobile wave solution,\cite{LO} in
contrast to Fulde~-~Ferrell running wave,\cite{FF} of the FFLO
problem.

Nonzero anomalous average $\langle
{\hat{c}}^{}_{{\bm{K}}^{}_j/2-{\bm{k}}\downarrow}
{\hat{c}}^{}_{{\bm{K}}^{}_j/2+{\bm{k}}\uparrow}\rangle $ appears
as a result of averaging of the product of two annihilation
fermion operators over the canonical ensemble in which total
particle number $N$ fluctuate with respect to certain mean value
$\bar{N}$.\cite{BCS} In such an ensemble, all of the states with
different $N$ close to $\bar{N}$ should be coherent so that pair
correlation function Eq.~(\ref{7}) describes ODLRO of
${\bm{K}}$~-~pairs in the SC condensate. Above $T^{}_c$, phase
coherence of the ground state turns out to be lost due to the fact
that ${\bm{K}}$-pairs in the states with different $N$ have got
random center-of-mass phases. Owing to above-mentioned instability
of the ground state with respect to a rise of QSS of
${\bm{K}}$-pair, relative motion phase of the wave function of
such a pair, included into coefficients ${\gamma}^{}_j$, can
remain locked up to temperatures far above $T^{}_c$.

One can think that a loss of relative-motion phase coherence with
heating might go through two steps. At first, $d$-wave
current-less superposition (\ref{7}) can be decomposed into two
orthogonal dimer superpositions with ${\gamma}^{}_1=\pm
{\gamma}^{}_3$, ${\gamma}^{}_2={\gamma}^{}_4=0$ and
${\gamma}^{}_2=\pm {\gamma}^{}_4$,
${\gamma}^{}_1={\gamma}^{}_3=0$. After that, at greater
temperature, dimer state can be disintegrated into free
${\bm{K}}$-pairs which survives up to their break at a temperature
that can be associated with the upper boundary of the PG state.
Temperature range, corresponding to lost center-of-mass phase
coherence but survived relative-motion phase coherence, can be
referred to the region of the PG state in which off-condensate SC
pairs can appear as spatially inhomogeneous ODSRO. We believe that
spatial pattern, like that observed by Kohsaka et
al.,\cite{Kohsaka} is described by current-less superpositions
Eq.~(\ref{7}) in which coefficients ${\gamma}^{}_j$ correspond to
random dimer configurations.

Recently, Berg et al.\cite{Berg} have considered dimer-like
(``striped'') ODLRO in the framework of the concept of SC pairing
with large momentum. Note that, as follows from numerical study of
Hubbard model on $4\times 4$ square lattice,\cite{Tsai} $d$-wave
checkerboard order as the ground state seems to be favorable with
respect to dimer-like one.

Momentum dependence of ODLRO parameter
${\Delta}^{}_{sc}({\bm{k}})$ determines the angle dependence of
the spectral weight, $W^{}_{CP}(\phi )$, of the SC coherent peak
appearing in the ARPES spectra below $T^{}_c$. In the case of
$d$-wave superconductor, it is tacitly assumed that SC order
parameter, taken on the FC, is proportional to ${\cos}{2{\phi}}$,
where Fermi angle $\phi$ is polar angle in the momentum space
counted from the antinodal direction. Therefore, $W^{}_{CP}(\phi
)$ should be a monotone function in the angle range $0\leq \phi
\leq {\pi}/4$ between the antinodal and nodal directions. However,
the ARPES study\cite{Kondo} shows unambiguously that the SC
spectral weight turns out to be highly non-monotonic: at first,
$W^{}_{CP}(\phi )$ increases from zero at ${\phi}={\pi}/4$ up to a
maximum at certain ${\phi}^{}_m$ and then exhibits a considerable
decrease if ${\phi}\rightarrow 0$. The spectral weight in the
antinodal region is strongly dependent on doping. Such a
non-monotonic behavior of $W^{}_{CP}(\phi )$ is
explained\cite{Kondo} by a competition between superconductivity
and an insulating state developing in the antinodal region with
pronounced nesting feature of the FC. The insulating state should
result in a depletion of the SC pairing channel and, in
consequence of a decrease of the SC order parameter, in a lowering
of the spectral weight of the SC coherent peak. It should be noted
that the spectral weight in the antinodal region, observed by
Kondo et al.,\cite{Kondo} is considerably greater than that
corresponding to simple ${\cos}\, 2{\phi}$ dependence as shown
schematically in Fig.~5a.

The coherent peak disappears in the PG state above $T^{}_c$ where
spectral weight $W^{}_{PG}(\phi)$ is zero in a broad angle range
that can be referred to the nodal region (Fig.~5b). In the
antinodal region, $W^{}_{PG}(\phi)$ increases rapidly up to a
maximum when ${\phi}\rightarrow 0$. Insulating order, which might
be invoked to explain both $W^{}_{CP}(\phi)$ and
$W^{}_{PG}(\phi)$, is not discovered for now. We believe that a
competition of such a hidden order\cite{CLMN} with
superconductivity is not the only qualitative explanation of
observed spectral properties in the antinodal region. We have
shown that ${\bm{K}}$~-~pairing concept\cite{BK} leads to a
consistent explanation of the origin of the SC and PG states:
${\bm{K}}$~-~pairing in the antinodal region gives rise to both
these states. A decrease of $W^{}_{CP}(\phi)$ at
${\phi}\rightarrow 0$ can be associated with non-trivial zero
lines of the SC order parameter ${\Delta}^{}_{sc}({\bm{k}})$ shown
in Fig.~3 that is can be explained in just the same way as a
decrease of $W^{}_{CP}(\phi)$ at ${\phi}\rightarrow {\pi}/4$ due
to $d$-wave node. It is evident that such angle dependence of the
spectral weight of the coherent peak, appearing in both nodal and
antinodal regions owing to different microscopic mechanisms of SC
pairing, can be considered as a direct indication of the biordered
SC state. On the contrary, since the PG state is associated with
QSS wave function, one can expect that $W^{}_{PG}(\phi)=0$ in the
nodal region whereas nonzero $W^{}_{PG}(\phi)$ in the antinodal
one is compared with $W^{}_{CP}(\phi)$ that appears there below
$T^{}_c$. Due to a random phase of the wave function of QSS and
corresponding gap function ${\Delta}^{}_{pg}({\bm{k}})$, nodal
lines of these functions, that could be apparent in the antinodal
region, cannot give a detectable contribution into a decrease of
the spectral weight at ${\phi}\rightarrow 0$. Expected angle
dependencies of $W^{}_{CP}(\phi)$ and $W^{}_{PG}(\phi)$ are shown
schematically in Fig.~5.

We believe the antinodal region with pronounced nesting of the FC
gives rise to ${\bm{K}}$-pairing whereas conventional pairing with
${\bm{K}}=0$ develops in the nodal region where the FC shows no
signs of nesting. Thus, we do not oppose ${\bm{K}}$-pairing with
the conventional pairing: these two SC pairing channels with
slightly overlapped domains of definition in the momentum space
form indivisible biordered SC state together. A passage from the
antinodal region into the nodal one is accompanied with a
redistribution of the spectral weight between these two pairing
channels.

\section{Quasiparticle interference}

A rise of the coherence in the system of the antinodal
${\bm{K}}$-pairs below $T^{}_c$ should inevitably lead to
interference effects inherent in the SC state. Bogoliubov QPI
appears due to mixing of quasipatricle states with high spectral
weight that results in a modulation of the local density of states
(LDOS) in the real space. Such states, at given quasiparticle
energy (\ref{B}), are disposed in vicinities of the points
corresponding to maximal curvature of the isoline
$E(k^{}_x,k^{}_y)=E={\text{const}}$. In the case of biordered SC
state, pairing with zero total momentum dominates
${\bm{K}}$-pairing in the nodal region, therefore,
${\eta}({\bm{k}})\equiv 0$ in this region due to the fact that
${\varepsilon}(-{\bm{k}})= {\varepsilon}({\bm{k}})$. Thus,
quasiparticle spectrum in the nodal region turns out to be fully
symmetrical with respect to the Fermi level.

\begin{figure}
\includegraphics[scale=.46]{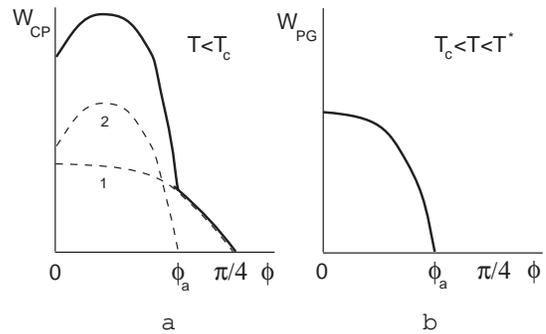}
\caption[*]{Angle dependence of the spectral weight
(schematically). {\bf{a}}. Spectral weight of the coherent SC peak
(dashed lines 1 and 2 correspond to $d$-wave order parameter
($\sim \cos{2{\phi}}$) and ${\bm{K}}$-pairing order parameter,
respectively. {\bf{b}}. Antinodal spectral weight of the PG state.
Here, ${\phi}^{}_a$ is a conditional angle boundary separating the
nodal and antinodal regions.}\label{FC.eps}
\end{figure}

On the contrary, ${\bm{K}}$-pairing dominates the pairing with
zero momentum in the antinodal region including near rectilinear
segments of the isolines in relatively small vicinity of the FC
which primarily forms the singularity in the self-consistency
equation. Because ${\varepsilon}({\bm{K}}/2+{\bm{k}})\approx
{\varepsilon}({\bm{K}}/2-{\bm{k}})$ in this vicinity,
quasiparticle spectrum in the antinodal region should be slightly
asymmetrical with respect to the Fermi level. Due to the fact that
there is a considerable increase of the deviation from mirror
nesting in a vicinity of conditional boundary separating nodal and
antinodal regions, one can expect a pronounced increase of the
asymmetry of quasiparticle spectrum in this vicinity. Such a
statement is compatible with photoemission data presented by Yang
et al.\cite{Yang} Also, it shows that an insulating state,
competing with the SC one, cannot be considered as the only origin
of the asymmetry observed by Yang et al.\cite{Yang} Indeed,
quasiparticle spectrum in the case of insulating state with gap
function $D({\bm{P}})$ has the form
\begin{equation}\label{A'}
E^{}_Q({\bm{p}})={\xi}({\bm{p}})\pm {\sqrt{{\eta}_{}^2({\bm{p}})+
D({\bm{p}})_{}^2}},
\end{equation}
where $2{\xi}({\bm{p}})\equiv
{\varepsilon}({\bm{p}})+{\varepsilon}({\bm{p}}+{\bm{Q}})$,
$2{\eta}({\bm{p}})\equiv
{\varepsilon}({\bm{p}})-{\varepsilon}({\bm{p}}+{\bm{Q}})$.
Therefore, imperfect nesting, that is a deviation from the nesting
condition (\ref{6}), just as imperfect mirror nesting, results in
the term (${\xi}({\bm{p}})$ or ${\eta}({\bm{p}})$ before the
square root) that originates electron-hole asymmetry in both
cases.

To study antinodal quasiparticle spectrum qualitatively, one can
neglect ${\eta}({\bm{k}})$ in Eq.~(\ref{A'}) owing to the fact
that antinodal segments of the FC appear as near rectilinear. For
this reason, at low quasiparticle energies, isoline shape can be
analyzed in general form. In such a case, isolines enclose the
gapless points of intersection of the FC and the nodal line. These
singular points can be found from equation system
${\xi}^{}_K(k^{}_x,k^{}_y)=0$, ${\Delta}^{}_K(k^{}_x,k^{}_y)=0$.

The nodal part of the FC can be approximated by an arc of a circle
whereas the nodal lines of $d$~-~wave superconductor are straight
lines $k^{}_y=\pm k^{}_x$. As a result, the quasiparticle isolines
become apparent as `banana'-like closed curves.\cite{McElroy}
Because Fermi energy ${\varepsilon}^{}_F$ exceeds $d$~-~wave SC
gap magnitude ${\Delta}^{}_m$ considerably, ${\varepsilon}^{}_F\gg
{\Delta}^{}_m$, `banana' turns out to be very thin so that exactly
its end points correspond to maximal curvature of the isoline.
This directly leads to the {\it{octet model}} of QPI in the nodal
region,\cite{McElroy} which defines a set of wave-vectors
${\bm{k}}^{}_i$ ($1\leq i\leq 8$), corresponding to such end
points, that determine LDOS pattern at given quasiparticle energy
$E$. The octet model is presented in Fig.~5a where we define the
main scattering momenta as ${\bm{q}}^{}_i=
{\bm{k}}^{}_i-{\bm{k}}^{}_1$. It should be noted that our
definition of ${\bm{q}}^{}_i$ is somewhat different from that
given by Kohsaka et el.\cite{Kohsaka} These two definitions are
mutually complementary in the reciprocal lattice.

\begin{figure}
\includegraphics[scale=.39]{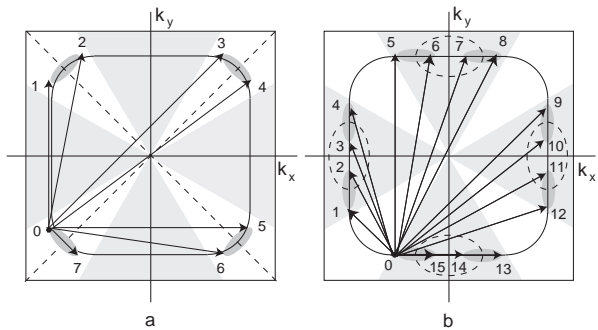}
\caption[*]{Main scattering momenta determining nodal ({\bf{a}},
in accordance with the ``octet'' model\cite{Kohsaka}) and
antinodal ({\bf{b}}) QPI pattern. FC and nodal line of the gap
function are presented as solid and dashed lines, respectively.
Small `bananas' are showed as shadowed ovals. Antinodal sectors of
the Brillouin zone are shadowed. Integers present the subscript
enumerating main scattering momenta.}\label{FC.eps}
\end{figure}

As one can see from Fig.~3, singular points of antinodal
quasiparticle spectrum $E({\bm{k}})$ are symmetrically, with
respect to the antinodal directions, disposed on near nested
pieces of the FC. To study a shape of the isolines, one can
approximate these pieces by straight lines and any of the nodal
lines in a small vicinity of the singular point by an arc of a
circle. Evidently, under the condition that ${\varepsilon}^{}_F\gg
{\Delta}^{}_m$, all isolines in this vicinity are `bananas'
enveloping a rectilinear part of the FC. Indeed, if $E\ll
{\varepsilon}^{}_F$, an isoline of the the quasiparticle
dispersion can be written in the form
\begin{equation}\label{10}
{\kappa}^2_x-{\kappa}^2_0=-{\kappa}^2_y\pm{\alpha}{\sqrt{{\kappa}^2_E
-{\kappa}^2_y}},
\end{equation}
where ${\kappa}^{}_x=k^{}_x/k^{}_F$, ${\kappa}^{}_y=k^{}_y/k^{}_F$
are dimensionless components of the relative motion momentum,
$k^{}_F$ is the Fermi momentum in the antinodal direction,
${\kappa}^{}_0$ is dimensionless radius of the nodal line,
${\kappa}^{}_E=E/2{\varepsilon}^{}_F$,
${\alpha}=2{\varepsilon}^{}_F/{\Delta}^{}_m(k^{}_Fa)_{}^2$. As
follows from Eq.~(\ref{10}), there are closed isolines only under
condition that $-{\kappa}^{}_E\leq {\kappa}^{}_y\leq
{\kappa}^{}_E$. Therefore, a transversal, with respect to the FC,
size of the isoline equals ${\kappa}^{}_t=2{\kappa}^{}_E\ll 1$. If
$E\rightarrow 0$, closed isolines shrink into two singular points
($\pm{\kappa}^{}_0,0$). A longitudinal size can be estimated as
${\kappa}^{}_l={\sqrt{{\kappa}^{2}_0+{\alpha}{\kappa}^{}_E}}-
{\sqrt{{\kappa}^{2}_0-{\alpha}{\kappa}^{}_E}}$, therefore, closed
isolines appear in quasiparticle energy range $0<E\leq
{\Delta}^{}_m(k^{}_0a)_{}^2$. In addition, one can examine that
${\kappa}^{}_t\ll{\kappa}^{}_l$. Energy $E^{}_m=
{\Delta}^{}_m(k^{}_0a)_{}^2$ corresponds to a topological
transition from closed, at $E<E^{}_m$, to opened, at $E>E^{}_m$,
isolines. Because opened isoline has no points of considerable
curvature, the topological transition should result in a
degradation of the interference pattern. Contrariwise, due to a
large curvature of the closed isoline in small vicinities of its
end points, exactly these vicinities should primarily contribute
into the QPI. Therefore, following McElroy et al.,\cite{McElroy}
one can introduce a set of momenta ${\bm{q}}^{}_i(E)$ connecting
different end points. Here, subscript $i$ runs from $1$ to $2n-1$
where $n$ is a number of singular points of quasiparticle
dispersion (\ref{A'}). Such main scattering momenta, defined as
${\bm{q}}^{}_i={\bm{k}}^{}_i-{\bm{k}}^{}_1$ for any $i\neq 1$,
should determine the real-space interference pattern. The pattern
caused by the antinodal QPI turns out to be considerably more
complicated in comparison with the nodal one even in the simplest
case corresponding to the only closed nodal line in each of four
crystal equivalent parts of the antinodal region as shown in
Fig.~6b. The full set of momenta ${\bm{q}}^{}_i$, following from
non-trivial momentum dependence of the gap function shown in
Fig.~3, should result in the real-space antinodal QPI pattern that
can be considered as originating from fairly uniform distribution
of scattering momenta. Therefore, it seems highly probable that
the antinodal QPI pattern should be considerably more smooth with
respect to the nodal one.

It is clear that due to expansion of the closed isolines
$E({\bm{k}})=E$ with an increase of $E$, there is a variation
(rotation and decrease or increase of the absolute value) of the
main scattering momenta, ${\bm{q}}^{}_i={\bm{q}}^{}_i(E)$ with
$E$. All of the nodal scattering momenta are dispersive, varying
with $E$ in accordance with the octet model.\cite{Kohsaka} On the
contrary, among the antinodal scattering momenta, there are some
non-dispersive, such as ``immobile'' ${\bm{q}}^{}_{5}$ and
${\bm{q}}^{}_{14}$ shown in Fig.~6b, that, owing to nesting
feature of the FC, remain independent of $E$ at small
quasiparticle energies. Strictly speaking, only such ``immobile''
${\bm{q}}^{}_i$ can contribute into checkerboard real-space
modulation in the SC state.\cite{Wang}

Nodes of the antinodal quasiparticle spectrum result in the fact
that, at a finite temperature, thermal equilibrium quasiparticles
are excited not only near $d$~-~wave nodes in the nodal
region\cite{Gedik} but in the antinodal one as well. Moreover, the
equilibrium population of the antinodal quasiparticles may
considerably exceed their population in vicinities of the
$d$~-~wave nodes. This may occur if line of zeroes of the
antinodal gap function, in contrast to the nodal one, turns out to
be close to the FC in its extended vicinity.

\section{Conclusion}

Our concept of ${\bm{K}}$-pairing in the cuprates is based on two
complementary statements: $1^{\circ}$ screened Coulomb repulsion
is the underlying SC pairing interaction; $2^{\circ}$ large
momentum of ${\bm{K}}$-pair arises due to nesting feature of the
FC. SC ${\bm{K}}$-pairing, prevailing in the antinodal region of
the momentum space, leads directly to uniform explanation of
spatial inhomogeneity of both SC state in the form of checkerboard
PDW and striped PG state formed by incoherent ${\bm{K}}$-pairs.
${\bm{K}}$-pairing, together with the conventional SC pairing with
zero momentum prevailing in the nodal region, results in an
indivisible biordered SC state which naturally explains the
peculiarities of the angle dependence of the spectral weight both
below and above $T^{}_c$. Complicated momentum dependence of the
gap function in the antinodal region should lead to fairly reach
antinodal QPI resulting in relatively smooth real-space
interference pattern.

\begin{acknowledgments}

This work was supported in part by the Russian Foundation for
Basic Research (Project Nos. 08-02-00490 and 09-02-00682).

\end{acknowledgments}

\end{document}